\documentclass[aps,showpacs,amssymb,amsfonts,twocolumn,pra]{revtex4}
\usepackage{amsmath,bbm,mathrsfs}
\usepackage{graphicx}
\usepackage{amsfonts}
\usepackage{amssymb}

\def\textbf#1{{\bf #1}}
\def\be{\begin{equation}}
\def\ee{\end{equation}}
\def\ben{\begin{eqnarray}}
\def\een{\end{eqnarray}}
\def\eea{\end{array}}
\def\bea{\begin{array}}
\newcommand{\ot}[0]{\otimes}

\newcommand{\bei}{\begin{itemize}}
\newcommand{\eei}{\end{itemize}}
\newcommand{\ket}[1]{|#1\rangle}
\newcommand{\bra}[1]{\langle#1|}

\newcommand{\proj}[1]{\ket{#1}\!\bra{#1}}
\newcommand{\braket}[2]{\langle{#1}|{#2}\rangle}

\def\blacksquare{\vrule height 4pt width 3pt depth2pt}

\begin{document}

\title{Searching for extremal PPT entangled states}

\author{Remigiusz Augusiak$^{1}$, Janusz Grabowski$^{2}$, Marek Ku\'s$^{3}$, and Maciej Lewenstein$^{1,4}$}
\affiliation{$^1$ICFO--Institut de Ci\`encies Fot\`oniques,
Parc Mediterrani de la Tecnologia, Castelldefels, 08860 Spain,\\
$^2$Faculty of Mathematics and Natural Sciences, College of Sciences,
Cardinal Stefan Wyszy\'nski University, Warszawa, Poland, \\
$^3$Center for Theoretical Physics, Polish Academy of Sciences, Warszawa, Poland,  \\
$^4$ICREA--Instituci\'o Catalana  de Recerca i Estudis Avan\c
cats, Barcelona, Spain}

\begin{abstract}
We study extremality in various sets of states that have positive
partial transposes. One of the tools we use for this purpose is
the recently formulated criterion allowing to judge if a given
state is extremal in the set of PPT states. First we investigate
qubit--ququart states and show that the only candidates for
extremal PPT entangled states (PPTES) have ranks of the state and
its partial transposition  $(5,5)$ or $(5,6)$ (equivalently
$(6,5)$). Then, examples of extremal states of $(5,5)$ type and
the so--called edge states of type $(5,6)$ are provided. We also
make an attempt to explore the set of PPT states with ranks
$(5,6)$. Finally, we discuss what are the possible configurations
of ranks of density matrices and their respective partial
transposition in general three-qubit and four-qubit symmetric
states for which there may exist extremal entangled PPT states.
For instance in the first case we show that the only possibilities
are $(4,4,4)$ and $(4,4,5)$.
\end{abstract}

\maketitle

\section{Introduction}

Entanglement is the property of states of composite quantum
systems which is the major resource for quantum information
\cite{rmp}. In the recent years considerable interest has been
devoted to the {\it entanglement problem}, i.e., the question of
determination whether a given state $\varrho$ of a composite
systems is entangled or not. This question, although it can be
formulated in an elementary way accessible to a fresh student of
physics, has very deep aspects and is related to applications
(quantum information, quantum communication, quantum metrology)
and advanced interdisciplinary fundamental problems. The latter
concern physics (properties of correlations in entangled states,
preparation, manipulation, and detection), mathematics (theory of
positive maps on $C^*$--algebras), philosophy (non--locality of
quantum mechanics), and computer science (quantum information).
Operational criteria for entanglement checking exist only in very
special cases. A famous Peres criterion says that every
non--entangled (separable) state has a positive partial transpose
(PPT) \cite{Peres} (see also Ref. \cite{Choi}). In two qubit and
qubit--qutrit systems the partial transposition gives necessary
and sufficient operational criterion of entanglement
\cite{MHPHRHPLA}. In all higher-dimensional systems there exist
PPT entangled states (PPTES, for the first examples see
Ref.~\cite{PHPLA}).

Recently, the interest in the entanglement problem has evidently become to
cease. The reasons for that are twofold: on the one hand Gurvits
\cite{Gurvits} has demonstrated that the entanglement problem is $NP$--hard,
i.e.\ finding operational criteria in higher dimensions is not likely. On the
other hand, for low dimensions Doherty {\it et al.} \cite{Doherty1} have
formulated a very efficient numerical test employing methods of
semi--definite programming and optimization.

Entanglement, quantum correlations and quantum measurement theory
were in the center of interests of the late Krzysztof
W\'odkiewicz, who made recently remarkable contributions to the
subject. His works concentrated on the entanglement and
correlations in finite--dimensional and continuous variable
systems \cite{KWcv}, quantum non-locality \cite{KWnl}, and
implementations in quantum optical systems \cite{KWim}. This
paper, in which we reconsider the entanglement problem, is
dedicated to his memory.

Our approach is not targeted at operational criteria, rather we
attempt to characterize and parameterize the whole set of extremal
PPTES, in particular for $2 \otimes 4$ systems. Parameterizing
this set allows one to construct and describe the set of necessary
and sufficient entanglement witnesses \cite{witness} that fully
characterize and detect all PPTES. For this purpose we need to
identify all the extremal entangled states in the set of PPT
states. In earlier works some of the authors have characterized
other class of PPT entangled states closely related to the
extremal ones. These are the so--called {\it edge states}
\cite{2xN,edge} (for some examples of the edge states see e.g.
\cite{PHPLA,UPB,Ha2,Lieven,Ha2007}). Quite obviously, edge states
are the only candidates for the extremal PPTES as every extremal
PPT entangled state is also of the edge type. In the search for
these extremal states we arrived at a simple criterion which
reduces the problem to the existence of solutions of a system of
linear equations. Later\footnote{After acceptance of the previous
version of the manuscript (see arXiv:0907.4979v2), J. M. Leinaas
brought our attention to to Ref. \cite{Leinaas}.} we learned that
the same result has already been worked out in Ref. \cite{Leinaas}
(see also Ref. \cite{Ha}). Also, in Ref. \cite{Leinaas} the
existence of extremal entangled PPTES of bi--rank $(5,6)$ (see
below for the explanation of the notation) was confirmed
numerically. In what follows we provide our formulation of the
criterion but one of the main purposes of the paper is to study
its applicability in various systems as qubit--ququart,
three--qubit, and four--qubit symmetric density matrices. We also
make a step towards understanding the structure of extremal states
in $2\ot 4$ systems.

It should be noticed that some effort has been recently devoted to
a search for extremal PPT entangled states in qutrit--qutrit
systems. In Ref. \cite{HaKyePark} $3\ot 3$ extremal PPT entangled
states of type $(4,4)$ were found. Then, the edge states provided in Refs.
\cite{Lieven,Ha2007} of type $(5,5)$ and $(6,6)$ were shown to be
extremal in \cite{Kim} and \cite{Ha}, respectively.

The paper is organized as follows. In Section~\ref{SecII} we
provide our formulation of the criterion. First, we discuss a
simple observation concerning extremal points of an intersection
of two convex sets. The observation is then applied to convex sets
of PPTES and the operational extremality criterion is formulated.
These ideas are then applied in Sec. \ref{SecIII} to the case of
$2\otimes 4$ systems. We discuss general properties of PPTES in
such systems and then focus our attention on states that have
ranks of the state and its partial transposition $(5,5)$ or
$(5,6)$ (equivalently $(6,5)$), and which are the only candidates
for extremal PPTES. In the case of bi--rank $(5,5)$ we prove that
every edge state is extremal in the set of PPT states and provide
examples of extremal states of the type $(5,5)$. Also, examples of
$(5,6)$ edge states are provided. In Sec. \ref{SecIV} we make an
attempt to explore the structure of PPTES of type $(5,6)$.
Finally, in Sec. \ref{SecV} we discuss the applicability of the
criterion in multipartite systems as three-qubit and four-qubit
symmetric states. We show also that any three-qubit edge state of
type $(4,4,4)$ is extremal and thus prove extremality of the bound
entangled state provided in Ref. \cite{UPB}.

Our results are primarily of fundamental
interest. They shed light on geometry of quantum states in general
\cite{zyczkos} and on the structure and nature of the very complex
convex set of PPTES, in $2\otimes 4$ systems in particular.

%
\section{Extremality criterion}
\label{SecII}
%
%

Here we present our formulation of the criterion
\cite{Leinaas} for judging if a given element of the set of PPTES
is extremal. For further benefits it is desirable to set first the
notation and explain in more details notions which have already
appeared in the Introduction.

\subsection{Preliminaries}

Let then $\varrho$ be a bipartite state acting on a product Hilbert space
$\mathcal{H}=\mathbb{C}^{d_{1}}\ot \mathbb{C}^{d_{2}}$. All such states
constitute a convex set which we denote by $\mathcal{D}_{d_{1},d_{2}}$. It
consists of two disjoint subsets of separable and entangled states. The
distinction between both sets is due to Werner \cite{Werner}. Following
\cite{Werner}, we call {\it separable} any state acting on $\mathcal{H}$ that
can be written as a convex combination of product states, that is
\begin{equation}\label{separable}
\varrho=\sum_{i}p_{i}\varrho_{1}^{(i)}\ot\varrho_{2}^{(i)},\qquad
p_{i}\geq 0,\qquad \sum_{i}p_{i}=1,
\end{equation}
with $\varrho_{j}^{(i)}$ acting on $\mathbb{C}^{d_{j}}$ $(j=1,2)$.
If $\varrho$ does not admit the above form we say that $\varrho$
is {\it entangled}. One of the most famous and important tests of
separability is based on the notion of transposition. Namely, it
was noticed by Peres \cite{Peres} that application of the map
$T\ot I$ (hereafter called {\it the partial transposition}), where
$I$ stands for the identity map, to any separable state gives
other separable state. Simultaneously, when applied to an
entangled state $\varrho$ the partial transposition can give a
matrix (hereafter denoted by $\varrho^{\Gamma_{A}}\equiv(T\ot
I)(\varrho)$ or $\varrho^{\Gamma_{B}}\equiv(I\ot T)(\varrho)$)
that is no longer positive. Thus, the transposition map may serve
as a good entanglement detector. In fact, it was shown in Ref.
\cite{MHPHRHPLA} that in qubit--qubit and qubit--qutrit systems it
detects all the entangled states. Interestingly, there exist
states which do not admit the form (\ref{separable}), but
nevertheless their partial transposition is positive (for the
first examples see Ref. \cite{PHPLA}). Entangled states that have
positive partial transpose cannot be distilled with local
operations and classical communication to a pure maximally
entangled states, and therefore are also called {\it bound
entangled} \cite{MHPHRH}.

Generally the partial transposition maps the set of density
matrices $\mathcal{D}_{d_{1},d_{2}}$ onto other set (containing
also non--positive matrices) $\mathcal{D}_{d_{1},d_{2}}^{\Gamma}$
which, as one easily verifies, is also convex. The intersection of
both sets
$\mathcal{D}_{d_{1},d_{2}}^{\mathrm{PPT}}=\mathcal{D}_{d_{1},d_{2}}\cap
\mathcal{D}_{d_{1},d_{2}}^{\Gamma}$, which of course is convex,
contains all the states with positive partial transpositions. Here
we arrive at an important notion in the nomenclature
connected to the set of PPT states. Namely, certain of the PPTES
that lie on the boundary of
$\mathcal{D}_{d_{1},d_{2}}^{\mathrm{PPT}}$ are called edge states
\cite{edge}. The formal definition says that given PPT entangled
state $\varrho\in\mathcal{D}_{d_{1},d_{2}}$ is an {\it edge state}
if for all $\epsilon>0$ and separable pure states
$\ket{e,f}\equiv\ket{e}\ket{f}\in\mathbb{C}^{d_{1}}\ot
\mathbb{C}^{d_{2}}$ it holds that $\varrho-\epsilon\proj{e,f}$ is
not positive or its partial transposition is not positive. From
this definition it follows immediately that any edge state is an
entangled state that does not contain in its range such a
separable vector $\ket{e,f}$ that $\ket{e^{*},f}$ ($\ket{e^{*}}$
denotes the complex conjugation of $\ket{e}$) belongs to the range
of its partial transposition.

We denote further by $R(X)$, $K(X)$, and $r(X)$ the range, kernel and the
dimension of $R(X)$ (that is the rank) of the matrix $X$, respectively.
Moreover, as the pair of numbers $(r(\varrho),r(\varrho^{\Gamma_{A}}))$
occurred to be useful in a classification of the edge states \cite{edge} and
then commonly utilized in the study of PPT states (see e.g. Refs.
\cite{edge,Ha,Kim,Ha2,Lieven}), hereafter we shall call it {\it bi--rank} of
$\varrho$.

%
\subsection{Extremal points of an intersection}
\label{SecIIA}
%

\begin{figure}
\begin{center}
\includegraphics[width=0.25\textwidth]{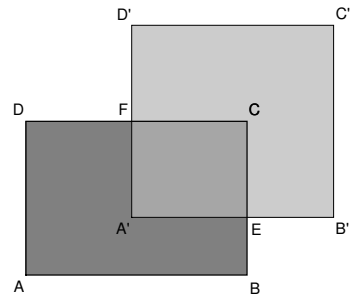}
\caption{ A simple example visualizing the observation formulated
in Sec. \ref{SecIIA}. The se of extremal points of the rectangle
$A'ECF$ resulting from an intersection of rectangles $ABCD$ and
$A'B'C'D'$ consist of some of the extremal points of both
rectangles or the ones that appear as intersections of their
respective edges. More precisely, the points $A'$ and $C$ are
extremal points of the rectangles $A'B'C'D$ and $ABCD$,
respectively. The Points $E$ and $F$ result from intersection of
the edges of these rectangles, namely, $A'B'$ and $BC$, and $CD$
and $D'A'$, respectively. Both the intersections give manifolds
consisting of a single point (the respective edges are
transversal). } \label{fig:intersection}
\end{center}
\end{figure}

Let us now discuss what are the extremal points of an intersection
of two nonempty convex sets $S_{1}$ and $S_{2}$ provided that
$S_{1}\cap S_{2}$ is nonempty. Recall that a given element $x\in
S$ is an {\it extremal element} of $S$ if it cannot be written as
a convex combination of elements from $S$ which are different from
$x$. There are two classes of extremal elements in $S_{1}\cap
S_{2}$ -- this statement we can formalize as the following (for an
illustrative example see Fig. \ref{fig:intersection}).

{\it Observation.} Let $S_{1}$ and $S_{2}$ be two convex sets.
Then the set of extremal points of $S_{1}\cap S_{2}$ consists of:
(i) the extremal points of $S_{1}$ and $S_{2}$ belonging to
$S_{1}\cap S_{2}$, (ii) possible new extremal points of $S_{1}\cap
S_{2}$ which belong to the intersection of boundaries of $S_{1}$
and $S_{2}$ and are not extremal points of $S_{1}$ and $S_{2}$.

To proceed in a more detailed way let us introduce some
definitions. Let $S$ be some convex set and let $x$, $y$ $(x\neq
y)$ be two elements from $S$. Then any subset of $S$ of the form
$[x,y]=\{\lambda x+(1-\lambda)y\,|\,x,y\in S,\;0\leq \lambda\leq
1\}$ is called {\it closed line segment} between $x$ and $y$.
Further, let $F$ be some subset of the convex set $S$. Then $F$ is
called {\it face} of $S$ if it is convex and the following
implication holds: if for any line segment $[x,y]\subseteq S$ such
that its interior point belongs to $F$ then $[x,y]\subseteq F$.
Equivalently, $F\subseteq S$ is a face of $S$ if from the fact
$x,y\in S$ and $(x+y)/2\in F$ it follows that $x,y\in F$. Thus,
$x\in S$ is an extremal point of $S$ if the one--element set
$\{x\}$ is a face of $S$.

\begin{figure}
\begin{center}
\includegraphics[width=0.5\textwidth]{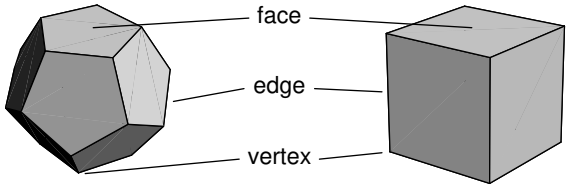}
\caption{The dodecahedron on the left and the known cube on the
right. A polyhedron consists of polygonal faces, their sides are
known as edges, and the corners as vertices. Edges and vertices
are also examples of faces of polyhedron.} \label{fig:faces}
\end{center}
\end{figure}

This definition formalizes the notion of a face of a convex
polygon or a convex polytope and generalizes it to an arbitrary
convex set. In this sense, vertices of a polytope are
zero--dimensional faces (extremal points), edges are
one--dimensional faces and the traditionally understood faces are
two--dimensional. The interior points form a two--dimensional face
(see Fig. \ref{fig:faces}). On the other hand, any point on the
boundary of a closed unit ball in $\mathbb{R}^n$, i.e., any point
of the unit $(n-1)$--dimensional sphere, is a face (and an extreme
point) of the ball.

Let us now discuss what are the main conclusions following the
above observation. Namely, if some convex sets $S_{1}$ and $S_{2}$
intersect in a point $x$ along an affine manifold of non--zero
dimension then $x$ cannot be extremal. Therefore $x$ belongs to
the second class of extremal points in $S_{1}\cap S_{2}$ if and
only if the intersection of the relevant faces is exactly $x$,
i.e., the faces are transversal at $x$ (see Fig.
\ref{fig:intersection}).

Let us also point out how the above observation can be utilized to
prove that a given $x$ is not an extremal element of $S_{1}\cap
S_{2}$. Namely, if $x$ belongs to the intersection of some faces
of $S_{1}$ and $S_{2}$, we can consider other element
$x\pm\epsilon y$ with $\epsilon$ denoting some arbitrarily small
positive real number and $y$ is an element of the affine space in
which $S_{1}$ and $S_{2}$ are convex subsets. If such element
belongs to the intersection of the mentioned faces the
intersection has a non--zero dimension and $x$ cannot be an
extremal element of $S_{1}\cap S_{2}$. The key point of this
remark is that to generate this manifold we do not have to
consider elements of the convex subsets $S_{1}$ and $S_{2}$. This,
as we will see in the next subsection, can be applied to convex
sets of density matrices with positive partial transposition.

\subsection{Criterion for extremality in convex sets of
quantum states with positive partial transposition} \label{criterion}

In the context of $\mathcal{D}_{d_{1},d_{2}}^{\mathrm{PPT}}$ the observation
of Sec.~\ref{SecIIA} says that the set of its extremal points is the sum of
two disjoint subsets. The first one consists of pure extremal states of
$\mathcal{D}_{d_{1},d_{2}}$ which are simultaneously extremal points of
$\mathcal{D}_{d_{1},d_{2}}^{\Gamma}$. The second subset consists of the
extremal points that could appear from intersection of boundaries of
$\mathcal{D}_{d_{1},d_{2}}$ and $\mathcal{D}_{d_{1},d_{2}}^{\Gamma}$ and are
not extremal points of both these sets. If a given $\varrho$ is neither an
extremal point of $\mathcal{D}_{d_{1},d_{2}}$ nor of
$\mathcal{D}_{d_{1},d_{2}}^{\Gamma}$ then we have to check (assuming
obviously that it is an edge state) if
the faces to which $\varrho$ and $\varrho^{\Gamma_{A}}$ belong are
transversal. It can be seen that this reduces to the problem of
solving a system of linear equations. For this purpose we utilize
the technique mentioned in the preceding section of adding or
subtracting small elements not necessarily belonging to
$\mathcal{D}_{d_{1},d_{2}}$ or
$\mathcal{D}_{d_{1},d_{2}}^{\Gamma}$. As we will see below for
this aim we can use Hermitian matrices $H$ such that
$R(H)\subseteq R(\varrho)$ and $R(H^{\Gamma_{A}})\subseteq
R(\varrho^{\Gamma_{A}})$. To proceed more formally let us notice
that, as we know from Ref.~\cite{JGMKGM}, the face of
$\mathcal{D}_{d_{1},d_{2}}$ at a given $\varrho$ (which is not of
full rank as only such density matrices constitute faces of
$\mathcal{D}_{d_{1},d_{2}}$) consists of matrices of dimensions
$d_{1}d_{2}\times d_{1}d_{2}$ with kernels coinciding with the
kernel of $\varrho$ (this holds also for partial transpositions of
density matrices provided they are positive). On the other hand,
all Hermitian matrices of rank $k$ constitute $k^{2}$ dimensional
subspace of the $(d_{1}d_{2})^{2}$--dimensional Hilbert space $V$
(equipped with e.g.\ the Hilbert--Schmidt scalar product) of all
$d_{1}d_{2}\times d_{1}d_{2}$ Hermitian matrices. To check if the
intersection $V_{1}\cap V_{2}$ of two subspaces $V_{1}$ and
$V_{2}$ of dimension $k^{2}$ and $l^{2}$  is nonempty, one has to
solve the system of $2(d_{1}d_{2})^{2}-k^{2}-l^{2}$ linear
equations for $(d_{1}d_{2})^{2} $ variables.  A nontrivial,
nonzero-dimensional, solutions exist always if $(d_{1}d_{2})^2<
k^{2}+l^{2}$ (in the case of density matrices, due to the
normalization, we add one to the left--hand side of this
inequality). If our density matrix $\varrho$ has a bi--rank
$(k,l)$ admitting the existence of such solutions, it cannot be
extremal in $\mathcal{D}_{d_{1},d_{2}}^{\mathrm{PPT}}$ -- one can
always find a Hermitian matrix $H\in V_{1}\cap V_{2}$ with
$R(H)\subseteq R(\varrho)$ and $R(H^{\Gamma_{A}})\subseteq
R(\varrho^{T_{A}})$ such that
$\widetilde{\varrho}(\epsilon)=\varrho\pm\epsilon H$ is positive
and PPT. What is important, for small enough $\epsilon>0$ the
matrix $\widetilde{\varrho}(\epsilon)$ has the same bi--rank as
the initial state $\varrho$. From the fact that the face of
$\mathcal{D}_{d_{1},d_{2}}$ at a given  $\varrho$ (which is not of
full rank as only such density matrices constitute faces of
$\mathcal{D}_{d_{1},d_{2}}$) consists of matrices of dimensions
$d_{1}d_{2}\times d_{1}d_{2}$ with kernels coinciding with the
kernel of $\varrho$ it follows that
$\widetilde{\varrho}(\epsilon)$ has to belong to the same
intersection as $\varrho$ and in such case $\varrho$ cannot be
extremal in $\mathcal{D}_{d_{1},d_{2}}^{\mathrm{PPT}}$. In this
way we finally arrive at the criterion.

{\it Criterion.} Let $\varrho\in \mathcal{D}_{d_{1},d_{2}}$ be PPTES with the
bi--rank $(k,l)$. If there exists a non-trivial, i.e.\ non-proportional to
$\varrho$, solution of the system of linear equations described above, i.e.\
equations determining  Hermitian matrices $H$ with $R(H)\subseteq R(\varrho)$
and $R(H^{\Gamma_{A}})\subseteq R(\varrho^{T_{A}})$), then  $\varrho$ is not
extremal in $\mathcal{D}_{d_{1},d_{2}}^{\mathrm{PPT}}$.

It should be emphasized that the above technique of adding and
subtracting Hermitian matrices and the criterion following it were
already provided in Ref. \cite{Leinaas}.

In conclusion, to prove that some
$\varrho$ is not extremal in the set of PPT states one can always
try to ``generate'' the nonzero-dimensional manifold by adding or
subtracting $\epsilon H$ with $H$ being a general Hermitian
matrix $H$ and $\epsilon$ some sufficiently small positive number.
This, however, has to be done in such a way that the
resulting state $\widetilde{\varrho}(\epsilon)=\varrho\pm\epsilon
H$, as well as its partial transposition have
the same kernels as, respectively, $\varrho$ and
$\varrho^{\Gamma_{A}}$. More precisely we need to have
$R(H)\subseteq R(\varrho)$ and $R(H^{\Gamma_{A}})\subseteq
R(\varrho^{\Gamma_{A}})$). In this way we remain on the same faces
of $\mathcal{D}_{d_{1},d_{2}}$ and
$\mathcal{D}_{d_{1},d_{2}}^{\Gamma}$ (as the partial transposition
of PPT state is still a legitimate state then the observation
above works also for partial transpositions of density matrices)
as $\varrho$ itself.

%
\section{The set of $2\ot 4$ states with the positive partial transposition}
\label{SecIII}
%
%
Here, we sketch shortly the overall picture of qubit--ququart
states with respect to the extremality in the set of PPT states.
First we recall literature results from which it follows that the
only possible cases of PPT entangled states that could be extremal
are the ones with bi--rank $(5,5)$ and $(5,6)$ (and equivalently
$(6,5)$), and $(6,6)$. Next, basing on the observation from
Sec.~\ref{SecII}, we will exclude the case of bi--rank $(6,6)$.
Then, we show that any edge state of bi--rank $(5,5)$ is extremal
in $\mathcal{D}_{2,4}^{\mathrm{PPT}}$ which implies that the
famous examples of $2\ot 4$ PPT entangled states provided in Ref.
\cite{PHPLA} are extremal. Finally, we present a two--parameter
class of edge states that depending on the parameters have
bi--rank $(5,6)$ or $(5,5)$, giving at the same time next example
of extremal $(5,5)$ states.

%
\subsection{The general situation}
\label{SecIIIA}
%
%

Let us here discuss shortly why the only possible cases in which there could
be PPT entangled extremal states are these of the bi--rank $(5,5)$, $(5,6)$
(equivalently $(6,5)$), and $(6,6)$. At the very beginning we can rule out
the cases in which either $\varrho$ or $\varrho^{\Gamma_{A}}$ are of full
rank since then the density matrix lie in the interior of the set of PPT
states (see e.g. Ref.~\cite{JGMKGM}).

To deal with the remaining cases we can assume that the given $\varrho$ is
supported on full $\mathbb{C}^{2}\ot\mathbb{C}^{4}$. Otherwise we would deal
with a state that is effectively qubit--qutrit state and one knows that in
this case there are no PPT entangled states \cite{MHPHRHPLA}. On the same
footing we can assume that there is no separable vector $\ket{e,f}$ in the
kernel of $\varrho$. Otherwise, it would mean, according to Ref. \cite{2xN},
that $\varrho$ could be written as a mixture of a projector onto the
separable vector $\ket{e^{\perp},h}$ (with $\braket{e^{\perp}}{e}=0$, and a
certain $\ket{h}$) and some PPT state supported on $\mathbb{C}^{2}\ot
\mathbb{C}^{3}$,  and thus separable. In conclusion, in such a case $\varrho$
could be written as a mixture of separable states and thus would be separable
itself.

We know from Ref.~\cite{2xN} that if $\varrho$ is supported on
$\mathbb{C}^{2}\ot\mathbb{C}^{N}$ and is entangled then $r(\varrho)>N$. This
means that in what follows we can assume that
$r(\varrho),r(\varrho^{\Gamma_{A}})>4$ as otherwise we would deal with a
separable state. Then, following Ref.~\cite{2xN}, we divide our
considerations into two cases,
\begin{description}
  \item[(i)] $r(\varrho)+r(\varrho^{\Gamma_{A}})\leq 12,$
  \item[(ii)] $r(\varrho)+r(\varrho^{\Gamma_{A}})> 12.$
\end{description}
In the second case it was shown in Ref.~\cite{2xN} that there
always exists a separable vector $\ket{e,f}$ in the range of
$\varrho$ for which $\ket{e^{*},f}\in R(\varrho^{\Gamma_{A}})$.
Consequently, we may subtract this vector from $\varrho$ obtaining
another (unnormalized) state
\begin{equation}\label{tilda}
\widetilde{\varrho}=\varrho-\eta\proj{e,f},
\end{equation}
where $\eta=\min\{\eta_{0},\overline{\eta}_{0}\}$ with $\eta_{0}$
and $\overline{\eta}_{0}$ given by $\eta_{0}=(\langle
e,f|\varrho^{-1}|e,f\rangle)^{-1}$ and
$\overline{\eta}_{0}=(\langle
e^{*},f|(\varrho^{\Gamma_{A}})^{-1}|e^{*},f\rangle)^{-1}.$ Here
$\varrho^{-1}$ denotes the so--called pseudoinverse of $\varrho$,
i.e., inverse of $\varrho$ only on the projectors corresponding to
its nonzero eigenvalues.

Normalizing $\widetilde{\varrho}$ we infer from Eq. (\ref{tilda}) that
$\varrho$ can be written as
\begin{equation}
\varrho=(1-\eta)\widetilde{\varrho}+\eta\proj{e,f},
\end{equation}
with $0<\eta<1$ (this is because both parameters $\eta_{0}$ and
$\overline{\eta}_{0}$ satisfy $0<\eta_{0},\overline{\eta}_{0}<1$). The
important fact here is that the above procedure of subtraction product
vectors preserves positivity of partial transposition of $\varrho$. More
precisely, if $\varrho^{\Gamma_{A}}\geq $ then also
$\widetilde{\varrho}^{\,\Gamma_{A}}\geq 0$. All these mean that if
$r(\varrho)+r(\varrho^{\Gamma_{A}})>12$ we are always able to write $\varrho$
as a convex combination of two states with positive partial transposition.
Thus in this case there are no PPT entangled extremal states in the convex
set $\mathcal{D}_{2,4}^{\mathrm{PPT}}$ (we see that even edge states do not
exist in these cases).

So far we have ruled out most of the cases with respect to the
bi--rank. What remains are the states satisfying $r(\varrho)>4$,
$r(\varrho^{\Gamma_{A}})>4$, and
$r(\varrho)+r(\varrho^{\Gamma_{A}})\leq 12$. In the cases of
$(5,7)$ and $(7,5)$ it was shown in Ref.~\cite{Normal} that there
exists a product vector $\ket{e,f} \in R(\varrho)$ such that
$\ket{e^{*},f}\in R(\varrho^{\Gamma_{A}})$. This, according to
what was said above, means that also among states with bi--rank
$(5,7)$ or $(7,5)$ one cannot find PPT entangled extremal states.
The case of bi--rank $(6,6)$ can be ruled out by the criterion.
Here $k=l=6$ and therefore one has a system of $56$ linear
homogenous equations for $63$ variables (one is subtracted due to
normalization). According to what was said in Sec.~\ref{criterion}
this means that all $(6,6)$ PPTES are not extremal as for any such
state there exist a Hermitian $H$ satisfying $R(H)\subseteq
R(\varrho)$ and $R(H^{\Gamma_{A}})\subseteq
R(\varrho^{\Gamma_{A}})$. Thus an intersection of the
corresponding faces of $\mathcal{D}_{2,4}$ and
$\mathcal{D}_{2,4}^{\Gamma}$ is thus a nonzero-dimensional
manifold.

%
\subsection{The case of bi--rank $(5,5)$}
%

The remaining cases in which we can look for the extremal PPT
states are $(5,5)$ and $(5,6)$. In the first case of the bi--rank
$(5,5)$ examples of extremal PPT entangled states can be provided.
For instance, we can show that the states found in Ref.
\cite{PHPLA},
\begin{equation}
\sigma(b)=\frac{1}{1+7b}
\left(
\begin{array}{cccccccc}
b & 0 & 0 & 0 & 0 & b & 0 & 0 \\
0 & b & 0 & 0 & 0 & 0 & b & 0 \\
0 & 0 & b & 0 & 0 & 0 & 0 & b \\
0 & 0 & 0 & b & 0 & 0 & 0 & 0 \\
0 & 0 & 0 & 0 & \frac{1}{2}(1+b) & 0 & 0 &
\frac{1}{2}\sqrt{1-b^{2}}\\
b & 0 & 0 & 0 & 0 & b & 0 & 0 \\
0 & b & 0 & 0 & 0 & 0 & b & 0 \\
0 & 0 & b & 0 & \frac{1}{2}\sqrt{1-b^{2}} & 0 & 0 &
\frac{1}{2}(1+b)
\end{array}
 \right),
\end{equation}
are extremal for $b\in(0,1)$ (for $b=0,1$ the state is separable).
For this aim let us prove the following theorem.

{\it Theorem.} Any $2\ot 4$ edge state with bi--rank $(5,5)$ is
extremal in the set of PPT states
$\mathcal{D}_{2,4}^{\mathrm{PPT}}$.

{\it Proof. (a.a.)} Let us consider an $2\ot 4$ edge state
$\sigma$ with the bi--rank $(5,5)$ and assume that it is not
extremal in $\mathcal{D}_{2,4}^{\mathrm{PPT}}$. This means that it
admits the form $\sigma=\lambda\sigma_{1}+(1-\lambda)\sigma_{2}$
with some $\lambda\in(0,1)$ and $\sigma_{i}\in
\mathcal{D}_{2,4}^{\mathrm{PPT}}$ $(i=1,2)$. Since $\sigma$ is
edge state it is clear that both the density matrices $\sigma_{i}$
have to be entangled. Also, it is easy to see that the conditions
$R(\sigma_{i})\subseteq R(\sigma)$ and
$R(\sigma_{i}^{\Gamma_{A}})\subseteq R(\sigma^{\Gamma_{A}})$ hold
for $i=1,2$. As a conclusion one of these density matrices, say
$\sigma_{1}$, can be subtracted from $\sigma$ with some small
portion $\epsilon>0$ in such a way that the resulting matrix, as
well as its partial transposition, remain positive. More
precisely, we can consider the state
$\widetilde{\sigma}=(\sigma-\epsilon\sigma_{1})/(1-\epsilon)$
which obviously has the rank $r(\widetilde{\sigma})\leq 5$. The
parameter $\epsilon$ can be set so that either
$r(\widetilde{\sigma})=4$ or $r(\widetilde{\sigma}^{\Gamma})=4$.
Consequently, by virtue of results of Ref.~\cite{2xN} where it was
shown that any PPT state supported on $2\ot 4$ with rank $4$ is
separable, the latter means that $\widetilde{\sigma}$ is
separable. As a result the initial state $\sigma$ can be written
in the form
\begin{equation}
\sigma=(1-\epsilon)\widetilde{\sigma}+\epsilon\sigma_{1},
\end{equation}
which means that $\sigma$ is a convex combination of a separable
and an entangled state. This, however, is in contradiction with
the assumption that $\sigma$ in an edge state. Thus $\sigma$ has
to be extremal what finishes the proof. $\blacksquare$

Notice that the above statement remains true if one relaxes the
condition of being an edge state to being a PPT entangled state.
That is, all PPT entangled states of bi-rank $(5,5)$ are extremal.

It was shown in Ref.~\cite{PHPLA} that $\sigma(b)$ for $b\in
(0,1)$ are edge states and therefore, due to the above theorem,
they are extremal in the set of $2\ot 4$ PPT states.

Let us remark that all the cases satisfying
$r(\varrho)+r(\varrho^{\Gamma_{A}})\geq 12$, as well as the case
of the bi-rank $(5,7)$ (and equivalently $(7,5)$) can be also
ruled out using the criterion.

\subsection{The case of bi--rank $(5,6)$}

The last case we need to deal with is the one with the bi--rank
$(5,6)$. It was numerically confirmed in Ref. \cite{Leinaas} that
there exist $(5,6)$ extremal PPT states in
$\mathcal{D}_{2,4}^{\mathrm{PPT}}$, however, explicit examples are
still missing.

In what follows let us consider the family of $2\ot 4$ PPT
entangled states with $r(\varrho)=5$ and
$r(\varrho^{\Gamma_{A}})=6$ which are edge but not extremal in the
set $\mathcal{D}_{2,4}^{\Gamma_{A}}$. The family is of the form
\begin{eqnarray}\label{przyklad}
\varrho(a,t)&=& \frac{1}{2(2+a+a^{-1})}\nonumber\\
&&\times\left(
\begin{array}{cccccccc}
a & 0 & 0 & 0 & 0 & -1 & 0 & 0 \\
0 & 1 & 0 & 0 & 0 & 0 & -1 & 0 \\
0 & 0 & a^{-1} & 0 & 0 & 0 & 0 & -1 \\
0 & 0 & 0 & 1 & t & 0 & 0 & 0 \\
0 & 0 & 0 & t & 1 & 0 & 0 & 0 \\
-1 & 0 & 0 & 0 & 0 & a^{-1} & 0 & 0 \\
0 & -1 & 0 & 0 & 0 & 0 & 1 & 0 \\
0 & 0 & -1 & 0 & 0 & 0 & 0 & a
\end{array}
\right).
\end{eqnarray}
The vector of eigenvalues of $\varrho(a,t)$ reads
\begin{eqnarray}
&&\hspace{-1cm}\lambda(\varrho(a,t))=\frac{1}{2(1+a)^{2}}\nonumber\\
&&\times\left(2a,1+a^{2},1+a^{2},a(1-t),a(1+t),0,0,0\right).
\end{eqnarray}
Thus, for $a>0$ and $-1<t<1$ the rank of $\varrho(a,t)$ is exactly five. Let
us now look at the partial transposition of $\varrho$, which is given by
\begin{eqnarray}\label{przykladTransp}
[\varrho(a,t)]^{\Gamma_{A}}&=&
\frac{1}{2(2+a+a^{-1})}\nonumber\\
&&\times\left(
\begin{array}{cccccccc}
a & 0 & 0 & 0 & 0 & 0 & 0 & t \\
0 & 1 & 0 & 0 & -1 & 0 & 0 & 0 \\
0 & 0 & a^{-1} & 0 & 0 & -1 & 0 & 0 \\
0 & 0 & 0 & 1 & 0 & 0 & -1 & 0 \\
0 & -1 & 0 & 0 & 1 & 0 & 0 & 0 \\
0 & 0 & -1 & 0 & 0 & a^{-1} & 0 & 0 \\
0 & 0 & 0 & -1 & 0 & 0 & 1 & 0 \\
t & 0 & 0 & 0 & 0 & 0 & 0 & a
\end{array}
\right).\nonumber\\
\end{eqnarray}
The vector of eigenvalues of $\varrho^{\Gamma_{A}}$ is of the form
\begin{eqnarray}\label{wlasneTr}
&&\hspace{-1cm}\lambda(\varrho(a,t)^{\Gamma_{A}})=\frac{1}{2(1+a)^{2}}\nonumber\\
&&\times\left(1-a,2a,2a,1,a(a-t),a(a+t),0,0\right).
\end{eqnarray}
For $a\geq  t$, $a+t\geq 0$, and $a<1$ one has
$[\varrho(a,t)]^{\Gamma_{A}}\geq 0$, hnece $\varrho(a,t)$ is a PPT state of
the bi--rank $(5,6)$ whenever $0<a<1$ and $|t|<a$. For $t=a$ it has the
bi--rank $(5,5)$ and is an edge state (one can check it using the method
presented in Ref.~\cite{PHPLA}). Due to the theorem from Sec.~\ref{SecIIIA}
it is also another example of an extremal state in
$\mathcal{D}_{2,4}^{\mathrm{PPT}}$.

To prove that the $\varrho(a,t)$ is entangled and thus bound
entangled we may use the already mentioned range criterion of Ref.
\cite{PHPLA}. After some algebra one finds that all the product
vectors in the range of $\varrho(a,t)$ can be written in the form
\begin{equation}\label{product}
\mathcal{A}(\alpha,t)(1; \alpha)\ot
(\alpha^{3};-\alpha^{2}/a;\alpha/a;1)\qquad
(\alpha\in\mathbb{C}\cup\{\infty\}),
\end{equation}
with $\mathcal{A}$ being some function of $\alpha$ and $t$. We used here the
fact that any vector from $\mathbb{C}^{2}$ can be written as
$\ket{0}+\alpha\ket{1}$ with $\alpha\in \mathbb{C}\cup\{\infty\}$ end
employed the convention $a\ket{0}+b\ket{1}\equiv(a;b)$ together
with a similar one for vectors in $\mathbb{C}^{4}$.

A similar analysis shows that the product vectors in the range of
$\varrho^{\Gamma_{A}}$ are of the form
\begin{eqnarray}\label{sepGamma}
(1;\alpha)&\!\!\!\ot\!\!\!& (\mathcal{B}(\alpha,a,t),-\alpha
\mathcal{B}(\alpha,a,t),\mathcal{C}(\alpha,a,t),-\alpha \mathcal{C}(\alpha,a,t)) \nonumber\\
&&\hspace{4cm}(\alpha\in\mathbb{C}\cup\{\infty\}),
\end{eqnarray}
with $\mathcal{B}$ and $\mathcal{C}$ denoting some functions. Simple calculations show
that one cannot find such a product vector $\ket{e,f}$ from
$R(\varrho(a,t))$ that $\ket{e^{*},f}\in
R(\varrho(a,t)^{\Gamma_{A}})$. More precisely, one cannot find
such functions $\mathcal{B}$ and $\mathcal{C}$ in Eq. (\ref{sepGamma}) that the vector
$\mathcal{A}(\alpha,t)(1; \alpha^{*})\ot
(\alpha^{3};-\alpha^{2}/a;\alpha/a;1)$ can be written in the above
form. Thus, for any $a$ and $t$ satisfying $0<a<1$ and $|t|<a$,
the matrix $\varrho(a,t)$ is a PPT entangled state. Moreover, the
same argument tell us that it is also an edge state.

Now, one immediately finds that $\varrho(a,t)$ cannot be extremal
in the set of $2\ot 4$ PPT states. This is because for a
fixed $a\in(0,1)$, $t\in (-a,a)$ can always be expressed as $t=(1-\lambda)t_{1}+\lambda
t_{2}$ with $\lambda\in(0,1)$ and $t_{i}\in (-a,a)$ $(i=1,2)$, which means that
for any $a\in(0,1)$ the state $\varrho(a,t)$ can be written as the following convex
combination
\begin{equation}
\varrho(a,t)=(1-\lambda)\varrho(a,t_{1})+\lambda\varrho(a,t_{2}).
\end{equation}

%
\section{Exploring PPT states of bi--rank $(5,6)$}
\label{SecIV}
%

We continue our considerations in the case of bi--rank $(5,6)$,
however, in the most general situation. As the discussion
presented in this section is a bit more mathematically demanding
it is helpful to outline first its contents before going into
details. The general aim of this section is to explore PPT states
with bi--rank $(5,6)$. It is stated in Ref. \cite{Leinaas} that
there exist extremal PPTES of this type, however, the general
structure of this class of states is not known and the main
purpose is to make a first attempt to understand it. For this
purpose, using the so--called canonical form of $2\ot N$ states,
we determine all the product vectors in ranges of $\varrho$ and
$\varrho^{\Gamma_{A}}$. Then, using the latter and following the
criterion, we propose for any $(5,6)$ PPT state such a Hermitian
matrix $H$ that $R(H)\subseteq R(\varrho)$ and look if
$R(H^{\Gamma_{A}})\subseteq R(\varrho^{\Gamma_{A}})$. This brings
the problem of extremality to the problem of solving a set of
linear equations. Although we cannot provide a solution to this
system in the general case, we can provide various sets of
conditions for $B$ (see Eq. \eqref{2x4a}) provided that the
solution has some particular form (and the corresponding state is
not extremal).

\subsection{Separable vectors in ranges of $\varrho$ and $\varrho^{\Gamma_{A}}$ in the case
of $(5,6)$} \label{SecIIIB}

Let us consider an arbitrary $2\ot 4$ state. It can be brought to
the so--called canonical form (see e.g.\
Ref.~\cite{Operational,Normal} and Ref. \cite{22N} in the context
of $2\ot 2\ot N$ states). To see it explicitly let us notice that
any such state has the form
\begin{equation}\label{2x4}
\varrho= \left(
\begin{array}{cc}
A & B\\
B^{\dagger} & C
\end{array}
\right),
\end{equation}
where $A$, $B$, and $C$ are some $4\times 4$ matrices and due to
the positivity of $\varrho$ one knows that $A$ and $C$ have to be
positive as well. Since we assume that $r(\varrho)=5$, we can also
assume that the matrix $C$ is of full rank. Otherwise, one sees to
see that the vector $\ket{1,f}$, where $\ket{f}\in K(C)$, belongs
to the kernel of $\varrho$. In view of what was said before this
means of course that $\varrho$ would be separable.

Thus, assuming that $r(C)=4$ we can bring $\varrho$ to its
canonical form by applying the transformation $\varrho\mapsto
(\mathbbm{1}_{2}\ot C^{-1/2})\varrho (\mathbbm{1}_{2}\ot
 C^{-1/2})$, which finally gives
\begin{equation}\label{form1}
\varrho'= \left(
\begin{array}{cc}
A' & B' \\
B'^{\dagger} & \mathbbm{1}_{4}
\end{array}
\right),
\end{equation}
where by $\mathbbm{1}_{d}$ we denoted $d\times d$ identity matrix.
It should be emphasized that the above transformation does not
change the rank, extremality as well as separability properties of
a given $\varrho$. Thus from our point of view both states
$\varrho$ and $(\mathbbm{1}_{2}\ot  C^{-1/2})\varrho
(\mathbbm{1}_{2}\ot  C^{-1/2})$ are completely equivalent.

We can simplify the form (\ref{form1}) even further. The positivity of
$\rho^\prime$ implies $A'=B'B'^{\dagger}+\Lambda$, where $\Lambda$ is some
positive matrix. Consequently, omitting all the primes we can write
\begin{equation}\label{2x4a}
\varrho= \left(
\begin{array}{cc}
BB^{\dagger}+\Lambda & B\\
B^{\dagger} & \mathbbm{1}_{4}
\end{array}
\right),
\end{equation}
where $B$ is some (in general non--Hermitian) matrix. Furthermore,
we can assume that $\det(B)=0$, which means that $r(B)=3$ and
$\det(B^{\dagger})=3$ and therefore there exist such $\ket{\phi}$
and $\ket{\widetilde{\phi}}$ from $\mathbb{C}^{4}$ that
$B\ket{\phi}=0$ and $B^{\dagger}\ket{\widetilde{\phi}}=0$,
respectively. This can be done by changing the basis in the first
subsystem. Precisely, we can apply the transformation
$U(\gamma,\delta)\ot\mathbbm{1}_{4}$ with $U(\gamma,\delta)$ given
by
\begin{equation}
U(\gamma,\delta)=
\left(
\begin{array}{cc}
\gamma & \delta \\
-\delta^{*} & \gamma^{*}
\end{array}
\right), \qquad (\gamma,\delta\in\mathbb{C}\setminus\{0\}),
\end{equation}
to $\varrho$. Then it suffices to solve the equation
$\det(\widetilde{B})=0$ which always has solutions. Finally,
$\Lambda$ is some positive matrix of rank one and thus can be
written as $\Lambda=\lambda_{1}\proj{\lambda_{1}}$.

Performing the partial transposition with respect
to the first subsystem we get
\begin{equation}\label{2x4Transp}
\varrho^{\,\Gamma_{A}}= \left(
\begin{array}{cc}
BB^{\dagger}+\Lambda & B^{\dagger}\\
B & \mathbbm{1}_{4}
\end{array}
\right)=\left(
\begin{array}{cc}
B^{\dagger}B+\widetilde{\Lambda} & B^{\dagger}\\
B & \mathbbm{1}_{4}
\end{array}
\right),
\end{equation}
with
$\widetilde{\Lambda}=[B,B^{\dagger}]+\Lambda=\lambda_{2}\proj{\lambda_{2}}+\lambda_{3}\proj{\lambda_{3}}$.
This form is similar to the one of $\varrho$ with the difference that
$\widetilde{\Lambda}$ is a combination of two projectors as we want to have
$r(\varrho^{\Gamma_{A}})=6$.

Let us now find the general form of separable vectors in the
ranges $\varrho$ and $\varrho^{\Gamma_{A}}$ and discuss their
properties. Due to the similarity of the forms of $\varrho$ and
$\varrho^{\Gamma_{A}}$ (see Eqs. (\ref{2x4a}) and
(\ref{2x4Transp})), it suffices to make the calculations for the
range of $\varrho$. As any (unnormalized) vector from
$\mathbb{C}^{2}\ot\mathbb{C}^{4}$ can be always written as
$\ket{0,g}+\ket{1,h}$, this can be done by solving the system of
equations $\varrho(\ket{0,g}+\ket{1,h})=\ket{e}\ket{f}$ with
$\ket{e}\in\mathbb{C}^{2}$ and $\ket{g}$, $\ket{h}$, and $\ket{f}$
denoting some vectors from $\mathbb{C}^{4}$. As previously we can
utilize the fact that any vector $\ket{e}$ from $\mathbb{C}^{2}$
can always be written as
$\ket{e}\equiv\ket{e(\alpha)}=\ket{0}+\alpha\ket{1}$ with
$\alpha\in\mathbb{C}\cup\{\infty\}$ (for the sake of simplicity we
forget about normalization, however we keep in mind that
$\ket{e(0)}=\ket{0}$ and $\ket{e(\infty)}=\ket{1}$). The resulting
set of equations reads
\begin{equation}
\left\{
\begin{array}{l}
(BB^{\dagger}+\Lambda)\ket{g}+B\ket{h}=\ket{f},\\[1ex]
B^{\dagger}\ket{g}+\ket{h}=\alpha\ket{f}.
\end{array}
\right.
\end{equation}
After some algebra one finds that the solution is of the form
\begin{equation}
\ket{f}\equiv a\ket{f(\alpha)}=a(\mathbbm{1}_{4}-\alpha
B)^{-1}\ket{\lambda_{1}},
\end{equation}
with $a=\lambda_{1}\langle \lambda_{1}|g\rangle$).  Except for the solutions
of $\det(\mathbbm{1}_{4}-\alpha B)=0$, the inverse of $\mathbbm{1}_{4}-\alpha
B$ exists for all $\alpha\in \mathbb{C}$.

The same reasoning in the case of $\varrho^{\Gamma_{A}}$ gives
\begin{equation}
\ket{\widetilde{f}(\alpha,b,c)}=(\mathbbm{1_{4}}-\alpha
B^{\dagger})^{-1}(b\ket{\lambda_{2}}+c\ket{\lambda_{3}}).
\end{equation}
with appropriate constants $b$ and $c$. Therefore the separable vectors from
$R(\varrho)$ and $R(\varrho^{T_{A}})$ are of the form
$a\ket{e(\alpha)}\ket{f(\alpha)}$ and
$\ket{e(\alpha)}\ket{\widetilde{f}(\alpha,b,c)}$, respectively. With some
additional effort we can determine a more detailed form of the separable
vectors in the range of $\varrho$ proving that the general form of
$\ket{f(\alpha)}$ is
\begin{equation}\label{wielomiany}
\ket{f(\alpha)}=\frac{1}{W(\alpha)}(W_{1}(\alpha);W_{2}(\alpha);W_{3}(\alpha);W_{4}(\alpha)),
\end{equation}
with $W(\alpha)$ and $W_{i}(\alpha)$ $(i=1,\ldots,4)$ being some polynomials
in $\alpha$ of degree at most three. To see this explicitly we notice that
since $r(\varrho)=5$ there exist three linearly independent vectors
$\ket{\phi_{i}}$ in $K(\varrho)$ orthogonal to
$\ket{e(\alpha)}\ket{f(\alpha)}$. Writing them in the form
\begin{equation}
\ket{\phi_{i}}=\ket{0}\ket{\widetilde{\phi}_{i}^{(0)}}
+\ket{1}\ket{\widetilde{\phi}_{i}^{(1)}},
\end{equation}
we get from the orthogonality conditions
\begin{equation}
\langle \widetilde{\phi}_{i}^{(0)}|f(\alpha)\rangle+\alpha \langle
\widetilde{\phi}_{i}^{(1)}|f(\alpha)\rangle=0, \qquad (i=1,2,3).
\end{equation}
By decomposing $\ket{f(\alpha)}=\sum_{k=1}^{4}f_{k}(\alpha)\ket{a_{k}}$ in
some basis $\{\ket{a_{k}}\}$ we get the following system of three homogenous
equations
\begin{equation}
\sum_{k}f_{k}(\alpha)\left(\langle
\widetilde{\phi}_{i}^{(0)}|a_{k}\rangle+\alpha \langle
\widetilde{\phi}_{i}^{(1)}|a_{k}\rangle\right)=0.
\end{equation}
We can always fix the value of one of $f_{k}$, e.g.\ by putting
$f_{4}(\alpha)=1$, and obtaining
\begin{eqnarray}
\label{dupa}
\sum_{k=1}^{3}f_{k}(\alpha)\left(\langle
\widetilde{\phi}_{i}^{(0)}|a_{k}\rangle+\alpha \langle
\widetilde{\phi}_{i}^{(1)}|a_{k}\rangle\right)&=&-\langle
\widetilde{\phi}_{i}^{(0)}|a_{4}\rangle\nonumber\\
&&-\alpha \langle
\widetilde{\phi}_{i}^{(1)}|a_{4}\rangle,\nonumber\\
\end{eqnarray}
for $i=1,2,3$. Now we deal with a system of three inhomogenous
equations (the inhomogeneity is nonzero as one can always find
such basis in $\mathbb{C}^{4}$ that at least one of the scalar
products on the right--hand side of (\ref{dupa}) is nonzero).
Solving the system we get the postulated form (\ref{wielomiany}).
Using a little bit more sophisticated reasoning one may also prove
that the polynomials $W_{i}$ $(i=1,2,3,4)$ are linearly
independent and therefore at least one of them must be of the
third degree. This also means that all the separable vectors in
$R(\varrho)$ can be brought to the form $(1;\alpha)\ot
(1;\alpha,\alpha^{2};\alpha^{3})$. Indeed, since $W_{i}$
$(i=1,2,3,4)$ are linearly independent there exist matrix $V$ with
$\det(V)\neq 0$ such that $\mathbbm{1}_{2}\ot
V\varrho\mathbbm{1}_{2}\ot V^{\dagger}$ has all the separable
vectors in its range of the above form.

\subsection{Density matrices of bi--rank $(5,6)$ that are not extremal}
Knowing the general form of product vectors from $R(\varrho)$ and
$R(\varrho^{\Gamma_{A}})$ of a general qubit--ququart state we can
now provide some conditions under which the $(5,6)$ PPT states are
not extremal. For this purpose let us consider the following
Hermitian operator
\begin{equation}
H=a_{1}a_{2}\ket{e(\alpha_{1})}\!\bra{e(\alpha_{2})}\ot
\ket{f(\alpha_{1})}\!\bra{f(\alpha_{2})}+h.c.
\end{equation}
with $\alpha_{1}\neq \alpha_{2}$. Its partial transposition with respect to
the first subsystem reads
\begin{equation}
H^{\Gamma_{A}}=a_{1}a_{2}\ket{e(\alpha_{2}^{*})}\!\bra{e(\alpha_{1}^{*})}\ot
\ket{f(\alpha_{1})}\!\bra{f(\alpha_{2})}+h.c.
\end{equation}
One sees that $R(H)\subseteq R(\varrho)$, still, however, we need
to check if $R(H^{\Gamma_{A}})\subseteq R(\varrho^{\Gamma_{A}})$.
The assumption $\alpha_{1}\neq \alpha_{2}$ is important since for
$\alpha_{1}=\alpha_{2}$ one could subtract a product vector from
$\varrho$ without spoiling its relevant properties, hence
$\varrho$ would not be an edge state.

We need to find such $\alpha_{1}$, $a_{1}$, $\alpha_{2}$, and
$a_{2}$ that
$a_{1}\ket{e(\alpha_{2})}\ket{f(\alpha_{1})}=\ket{e(\alpha_{2})}
\ket{\widetilde{f}(\alpha_{2}^{*},a_{1},b_{1})}$ and
$a_{2}\ket{e(\alpha^{*}_{1})}\ket{f(\alpha_{2})}=
\ket{e(\alpha_{1}^{*})}\ket{\widetilde{f}(\alpha_{1}^{*},b_{2},c_{2})}$
as both these vectors belong to the range of
$\varrho^{\Gamma_{A}}$. These conditions simplify to
$a_{1}\ket{f(\alpha_{1})}=\ket{\widetilde{f}(\alpha_{2}^{*},b_{1},c_{1})}$
and
$a_{2}\ket{f(\alpha_{2})}=\ket{\widetilde{f}(\alpha_{1}^{*},b_{2},c_{2})}$,
respectively which, in turn, can be brought to a system of linear
equations
\begin{equation}\label{Eq1}
(I-\alpha_{1} B)\ket{x}=a_{1}\ket{\lambda_{1}},
\end{equation}
\begin{equation}\label{Eq2}
(I-\alpha_{2}^{*}
B)\ket{x}=b_{1}\ket{\lambda_{2}}+c_{1}\ket{\lambda_{3}},
\end{equation}
where $\ket{x}$ is a vector from $\mathbb{C}^{4}$ that has to be determined.
We can introduce three vectors $\ket{f_{i}}$ $(i=1,2,3)$ orthogonal to
$\ket{\lambda_{1}}$ and two vectors $\ket{v_{i}}$ $(i=1,2)$ orthogonal to
$\ket{\lambda_{2}}$ and $\ket{\lambda_{3}}$. Projection of Eqs.~(\ref{Eq1})
and (\ref{Eq2}) onto the vectors $\ket{f_{i}}$ $(i=1,2,3)$ and
$\ket{\lambda_{1}}$ leads to the following systems of equations
\begin{equation}\label{uklad11}
\left\{
\begin{array}{l}
\bra{f_{i}}I-\alpha_{1}B\ket{x}=0, \qquad i=1,2,3,\\[1ex]
\bra{\lambda_{1}}I-\alpha_{1}B\ket{x}=a_{1},
\end{array}
\right.
\end{equation}
and onto the vectors $\ket{v_{i}}$ $(i=1,2)$ and $\ket{\lambda_{2}}$, and
$\ket{\lambda_{3}}$ to
\begin{equation}\label{uklad12}
\left\{
\begin{array}{l}
\bra{v_{i}}I-\alpha_{2}^{*}B^{\dagger}\ket{x}=0, \qquad i=1,2,\\[1ex]
\bra{\lambda_{2}}I-\alpha_{2}^{*}B^{\dagger}\ket{x}=
b_{1}+c_{1}\langle\lambda_{2}|\lambda_{3}\rangle,\\[1ex]
\bra{\lambda_{3}}I-\alpha_{2}^{*}B^{\dagger}\ket{x}
=b_{1}\langle\lambda_{3}|\lambda_{2}\rangle+c_{1}.
\end{array}
\right.
\end{equation}
The same procedure applied to the equation
$a_{2}\ket{f(\alpha_{2})}=\ket{\widetilde{f}(\alpha_{1}^{*},b_{2},c_{2})}$
leads to an analogous system of equations
\begin{equation}\label{uklad21}
\left\{
\begin{array}{l}
\bra{f_{i}}I-\alpha_{2}B\ket{y}=0, \qquad i=1,2,3,\\[1ex]
\bra{\lambda_{1}}I-\alpha_{2}B\ket{y}=a_{2},
\end{array}
\right.
\end{equation}
\begin{equation}\label{uklad22}
\left\{
\begin{array}{l}
\bra{v_{i}}I-\alpha_{1}^{*}B^{\dagger}\ket{y}=0, \qquad i=1,2,\\[1ex]
\bra{\lambda_{2}}I-\alpha_{1}^{*}B^{\dagger}\ket{y}
=b_{2}+c_{2}\langle\lambda_{2}|\lambda_{3}\rangle,\\[1ex]
\bra{\lambda_{3}}I-\alpha_{1}^{*}B^{\dagger}\ket{y}
=b_{2}\langle\lambda_{3}|\lambda_{2}\rangle+c_{2}.
\end{array}
\right.
\end{equation}
Together, the homogenous equations in (\ref{uklad11}),
(\ref{uklad12}), (\ref{uklad21}), and (\ref{uklad22}) give ten
equations for eight unknowns variables $x_{i}$ and $y_{i}$
$(i=1\ldots,4)$ being coordinates of $\ket{x}$ and $\ket{y}$ in
some basis in $\mathbb{C}^{4}$. From the remaining inhomogenous
equations one could determine $a_{i}$, $b_{i}$, and $c_{i}$
$(i=1,2)$. The most natural and, in our opinion, easiest way to
solve all the homogenous equations is to consider them separately
for $\ket{x}$ and $\ket{y}$. For this aim let us rewrite the
homogenous equations of \eqref{uklad11} and \eqref{uklad12} as
\begin{equation}
\label{dupa1}
\left\{
\begin{array}{ll}
\bra{f_{i}}I-\alpha_{1}B\ket{x}=0, \qquad i=1,2,3,\\[1ex]
\bra{v_{i}}I-\alpha_{2}^{*}B^{\dagger}\ket{x}=0, \qquad i=1,2.
\end{array}
\right.
\end{equation}
They constitute five linear equations for $\ket{x}$, which implies
that two determinants have to vanish; we take them to involve the
first three equations, together with one of the remaining two. In
effect each determinant has a form of a third order polynomial in
$\alpha_1$ and first order in $\alpha_2^*$. Such a system of two
polynomials has several solutions for the pairs
$(\alpha_1,\alpha_2)$ with $\alpha_1\ne \alpha_2$ (if there were a
solution $\alpha_1=\alpha_2$, we could subtract the projector on
the corresponding product vector from the state $\varrho$ and the
partially transposed one from the partially transposed state
keeping the positivity of both, which is incompatible with
$\varrho$ being an edge state). The remaining equations,
$\bra{f_{i}}I-\alpha_{2}B\ket{y}=0$, with  $i=1,2,3$ and
$\bra{v_{i}}I-\alpha_{1}^{*}B^{\dagger}\ket{y}=0$, with $i=1,2$ we
obtain by replacing in (\ref{dupa1}) $\ket{x}\to \ket{y}$, and
$\alpha_1\to\alpha_2$. Concluding, there always exist such pairs
of $\alpha_{1}$ and $\alpha_{2}$ that the set \eqref{dupa1} has a
nontrivial solution. Now, if the analogous set of equations for
the vector $\ket{y}$ has a solution for the particular
$\alpha_{1}$ and $\alpha_{2}$ solving \eqref{dupa1}, but
interchanged $\alpha_{1}\leftrightarrow \alpha_{2}$, then we have
a solution to the initial problem. We cannot find the general
solution to this problem and due to results of Ref. \cite{Leinaas}
it is even impossible. Thus, we consider some particular cases
with respect to $\alpha_{1}$ and $\alpha_{2}$, and assuming that
the solution is of some particular form we derive certain
conditions that have to be satisfied by the matrix $B$ (see
Eq.~\eqref{2x4}).

For $\alpha_{1}=\infty$ and $\alpha_{2}=0$ the homogenous
equations from \eqref{uklad11}, \eqref{uklad12}, \eqref{uklad21},
and \eqref{uklad22} reduce to
\begin{equation}\label{case11}
\left\{
\begin{array}{ll}
\bra{f_{i}}B\ket{x}=0, &\quad i=1,2,3,\\[1ex]
\bra{v_{i}}x\rangle=0, &\quad i=1,2,
\end{array}
\right.
\end{equation}
and
\begin{equation}\label{case11a}
\left\{
\begin{array}{ll}
\bra{f_{i}}y\rangle=0, &\quad i=1,2,3\\[1ex]
\bra{v_{i}}B^{\dagger}\ket{y}=0, &\quad i=1,2.
\end{array}
\right.
\end{equation}
From the first set we see that
$\ket{x}=\xi\ket{\lambda_{2}}+\eta\ket{\lambda_{3}}$, while from
the second one that $\ket{y}=\mu\ket{\lambda_{1}}$. So in the case
of $\alpha_{1}=\infty$ and $\alpha_{2}=0$ we are able to find
$\ket{x}$ and $\ket{y}$, however the matrix $B$ has to satisfy the
conditions
$\bra{f_{i}}B(\xi\ket{\lambda_{2}}+\eta\ket{\lambda_{3}})=0$ with
$i=1,2,3$ and $\bra{v_{i}}B^{\dagger}\ket{\lambda_{1}}=0$ for
$i=1,2$.

In the case $\alpha_{1}=0$ and $0\neq\alpha_{2}\neq\infty$ we have
\begin{equation}\label{case21}
\left\{
\begin{array}{ll}
\bra{f_{i}}x\rangle=0, &\quad i=1,2,3,\\[0.5ex]
\bra{v_{i}}I-\alpha_{2}^{*}B^{\dagger}\ket{x}=0, &\quad i=1,2,
\end{array}
\right.
\end{equation}
and
\begin{equation}\label{case22}
\left\{
\begin{array}{ll}
\bra{f_{i}}I-\alpha_{2}B\ket{y}=0, &\quad i=1,2,3,\\[0.5ex]
\bra{v_{i}}y\rangle=0, &\quad i=1,2.
\end{array}
\right.
\end{equation}
Here, analogously to the previous case, one has
$\ket{x}=\widetilde{\mu}\ket{\lambda_{1}}$ and
$\ket{y}=\widetilde{\xi}\ket{\lambda_{2}}+\widetilde{\eta}\ket{\lambda_{3}}$.
Inserting the above forms of $\ket{x}$ and $\ket{y}$ to Eqs. (\ref{case21})
and (\ref{case22}) we obtain five conditions for the matrix $B$. More
precisely, we have
$\bra{v_{i}}I-\alpha_{2}^{*}B^{\dagger}\ket{\lambda_{1}}=0$ with $i=1,2$ and
$\bra{f_{i}}I-\alpha_{2}B(\widetilde{\xi}\ket{\lambda_{2}}+\widetilde{\eta}\ket{\lambda_{3}})=0$
for $i=1,2,3$.

Finally in the case of $\alpha_{1}=\infty$ and $0\neq \alpha_{2}\neq \infty$
the systems of equations reads
\begin{equation}\label{sys1Homa}
\left\{
\begin{array}{ll}
\bra{f_{i}}B\ket{x}=0, &\quad i=1,2,3,\\[1ex]
\bra{v_{i}}I-\alpha_{2}^{*}B^{\dagger}\ket{x}=0, &\quad i=1,2,
\end{array}
\right.
\end{equation}
and
\begin{equation}\label{sys2Homa}
\left\{
\begin{array}{ll}
\bra{f_{i}}I-\alpha_{2}B\ket{y}=0, &\quad i=1,2,3,\\[1ex]
\bra{v_{i}}B^{\dagger}\ket{y}=0, &\quad i=1,2.
\end{array}
\right.
\end{equation}
All the provided equations in each of the above cases are
obviously equipped with conditions of vanishing determinants.

%
\section{Multi--qubit PPT extremal states}
\label{SecV}

It seems interesting to extend the criterion to the case of
many--qubits states. In what follows we discuss how it can be
applied to the general three and symmetric four qubit states with
all partial transposes positive (hereafter PPT states). As we will
see in both cases we deal with an intersection of three convex
sets. Let us then discuss how extremal points appear when three
convex sets $S_{1}$, $S_{2}$, and $S_{3}$ intersect provided that
$S_{1}\cap S_{2}\cap S_{3}$ is not an empty set. Following the
observation formulated in the previous sections (we can apply it
recursively to the pair of sets $S_{1}$ and $S_{2}$, and then to
the pair $S_{1}\cap S_{2}$, and $S_{3}$) one sees that in general
the set of extremal points of $S_{1}\cap S_{2}\cap S_{3}$ consists
of (i) extremal points of one of these three sets, (ii) points
that appear as an intersection of faces of {\it two} of these
three sets (but are elements of the interior of the remaining set)
and are not extremal in any of them, (iii) points that appear as
an intersection of faces of all the sets.

\subsection{Three qubits}

At the beginning let us notice that separability properties in
three-qubit systems were thoroughly studied in Ref. \cite{22N}.
Then all these states were classified in Ref. \cite{3qu}.

Let now $\mathcal{D}^{\mathrm{PPT}}_{2,2,2}$ be the set (obviously
convex) of all three--qubit states having all partial transposes
positive. 
It is clear that positivity of all partial transposes is
equivalent to positivity of arbitrary two elementary subsystems,
say $A$ and $B$ (positivity of the remaining partial transposes
$\varrho^{\Gamma_{C}}$, $\varrho^{\Gamma_{AC}}$, and $\varrho^{\Gamma_{BC}}$
is then guaranteed {\it via } the positivity of the global transpose).
Consequently, $\mathcal{D}^{\mathrm{PPT}}_{2,2,2}$ is intersection
of three convex sets, namely, the set of all three--qubit density
matrices denoted by $\mathcal{D}_{2,2,2}$ and two sets
$\mathcal{D}_{2,2,2}^{\Gamma_{A}}$ and
$\mathcal{D}_{2,2,2}^{\Gamma_{B}}$ containing partially transposed
three--qubit density matrices with respect to the $A$ and $B$
subsystem, respectively.

By $k$, $l$, and $m$ we denote ranks of $\varrho$,
$\varrho^{\Gamma_{A}}$, and $\varrho^{\Gamma_{B}}$, respectively
(generically $1\leq k,l,m\leq 8$, however, we can exclude the rank
one cases as they are already extremal). First, let us recall that
it was shown in Ref. \cite{22N} that every three--qubit state
$\varrho$ with $r(\varrho)\leq 3$ supported on
$\mathbb{C}^{2}\ot\mathbb{C}^{2}\ot\mathbb{C}^{2}$ is fully
separable. Thus we can assume that
$r(\varrho),r(\varrho^{\Gamma_{A}}),r(\varrho^{\Gamma_{B}})\geq 4$
as the same arguments can be applied to $\varrho^{\Gamma_{A}}$ and
$\varrho^{\Gamma_{B}}$. Then, it should be noticed that any
three-qubit state can be treated as $2\ot 4$ state with respect to
the partitions $A|BC$, $B|AC$, and $C|AB$. This, due to results of
Ref. \cite{2xN} means that every three-qubit PPT state with rank
$r(\varrho)=4$ is separable across all the above cuts. However, an
example of a state separable across $A|BC$, $B|AC$, and $C|AB$
cuts but nevertheless entangled was provided in Ref. \cite{UPB}
(the bound entanglement constructed from the so--called
unextendible product basis), which as we will see later is
extremal in $\mathcal{D}_{2,2,2}^{\mathrm{PPT}}$. Thus in what
follows we can put $k,l,m\geq 4$.

Now, we can follow analogous reasoning as in the bipartite case
and obtain the inequality $k^{2}+l^{2}+m^{2}>65$, which if
satisfied implies existence of such a Hermitian matrix $H$ that
$\widetilde{\varrho}(\epsilon)=\varrho\pm\epsilon H$ is a density
matrix for some nonzero $\epsilon$ with all partial transposes
positive and the same ranks $k,l,m$ (and thus $\varrho$ of such
ranks cannot be extremal). Below we consider only such triples
that satisfy $k\leq l\leq m$ as the remaining ones one obtains by
permuting the ranks $k,l,m$. It follows immediately from this
inequality that there do not exist PPT extremal states of the
second kind (see the discussion at the beginning of Sec.
\ref{SecV}). This is because if one of the ranks is maximal the
inequality becomes $i^{2}+j^{2}>1$ $(i,j=k,l,m)$ and is always
satisfied. On the other hand, application of this inequality to
the cases when all ranks are not maximal allows to exclude the
following cases $(4,4,6)$, $(4,4,7)$, $(4,5,5)$, $(4,5,6)$,
$(4,5,7)$, $(4,6,6)$, $(4,6,7)$, and $(4,7,7)$. As a result the
only cases that remain are $(4,4,4)$ and $(4,4,5)$. In the first
case an example of extremal PPTES was given in \cite{UPB}. To see
it explicitly one may prove a similar statement to the theorem
from Sec. \ref{SecIIIB}. Namely, adapting its proof and utilizing
the results of Ref. \cite{22N} we get the following theorem.

{\it Theorem 2.} Any three-qubit PPT entangled state supported on
$\mathbb{C}^{2}\ot\mathbb{C}^{2}\ot\mathbb{C}^{2}$ with ranks
$r(\varrho)=r(\varrho^{\Gamma_{A}})=r(\varrho^{\Gamma_{B}})=4$ is
extremal in $\mathcal{D}_{2,2,2}^{\mathrm{PPT}}$.

Since, by construction the three-qubit bound entangled state from
Ref. \cite{UPB} is an edge state it has to be extremal in
$\mathcal{D}_{2,2,2}^{\mathrm{PPT}}$ of type $(4,4,4)$.

\subsection{Four qubits in a symmetric state}

In the case of four qubits we restrict our discussion only to the
states acting on $\mathcal{S}((\mathbb{C}^{2})^{\ot 4})$, which is
a subspace of $(\mathbb{C}^{2})^{\ot 4}$ consisting of all pure
states symmetric under permutation of any subset of parties (let
us call such states symmetric). General $N$-qubit symmetric states
were recently investigated in numerous papers (see e.g. Refs.
\cite{Eckert,Korbicz,UshaDevi,Guhne}).

One of the motivations to study four-qubit symmetric states comes
from a long--standing open question whether there exist such
states with positive partial transpose with respect to all
subsystems. If it were so there would also exist extremal PPT
symmetric states. Notice that such states, however, with higher
number of qubits were found very recently in Ref. \cite{Guhne}. In
what follows we would like to discuss what are the possible cases
with respect to ranks of respective partial transposes for which
there could exist entangled PPT extremal symmetric states.

Notice relaxing the PPT condition to two--party transposes only as
for instance $\varrho^{\Gamma_{AB}}\geq 0$ and leaving the
possibility that single--particle partial transposes are
nonpositive, it is possible to find four--partite bound entangled
symmetric state \cite{Smolin} which even allows for maximal
violation of Bell inequality \cite{RAPH}. Interestingly, this
state was very recently realized experimentally \cite{Amselem}.

Let us discuss firstly some of the properties of mixed states
acting on $\mathcal{S}((\mathbb{C}^{2})^{\ot 4})$. Clearly, it is
spanned by the vectors $\ket{0}^{\ot 4}$, $\ket{1}^{\ot 4}$,
$\ket{W}$, $\ket{\overline{W}}$, and
$(1/\sqrt{6})(\ket{0011}+\ket{0101}+\ket{1001}+\ket{1010}+\ket{1100}+\ket{0110}),$
where $\ket{W}$ is the so--called $W$ state
$\ket{W}=(1/2)(\ket{0001}+\ket{0010}+\ket{0100}+\ket{1000})$ and
$\ket{\overline{W}}$ is obtained from $\ket{W}$ by replacing zeros
with ones and {\it vice versa}. Consequently, all symmetric
four--qubit density matrices have ranks at most five. It also
follows from the above that there are only two relevant partial
transposes defining set of PPT symmetric states. Namely,
positivity of partial transpose with respect to a single subsystem
(two subsystems) means positivity of partial transposes with
respect to all single subsystems $A,B,C,D$ (arbitrary pairs of
subsystems $AB$, $AC$, etc.). The three--party partial transposes
are equivalent to the single--party partial transposes {\it via}
the global transposition. Hence, as in the case of three--qubits,
the set of PPT symmetric states is an intersection of three sets,
namely, the set of symmetric density matrices and the sets of
partial transpositions of the latter with respect to the $A$ and
$AB$ subsystem, respectively. So, there are three relevant
parameters, $r(\varrho)$, $r(\varrho^{\Gamma_{A}})$, and
$r(\varrho^{\Gamma_{AB}})$. Using the fact that we deal with the
symmetric states we can easily impose bounds on these three ranks.
We already know that $r(\varrho)\leq 5$. Then, since
$\mathcal{S}((\mathbb{C}^{2})^{\ot 4})$ is a subspace of
$\mathbb{C}^{2}\ot\mathbb{C}^{4}$ (the $A$ subsystem acts on
$\mathbb{C}^{2}$ and the remaining $BCD$ subsystems act on
$\mathbb{C}^{4}$) it holds that $r(\varrho^{\Gamma_{A}})\leq 8$.
On the other hand, $\mathcal{S}((\mathbb{C}^{2})^{\ot 4})$ is also
a subspace of $\mathbb{C}^{3}\ot\mathbb{C}^{3}$ (the two--party
subsystems $AB$ and $CD$ act on $\mathbb{C}^{3}$) and hence
$r(\varrho^{\Gamma_{AB}})\leq 9$.

Further constraints on the ranks can be imposed by utilizing
results of Refs. \cite{2xN} and \cite{Eckert}. First of all from
Ref. \cite{Eckert} we know that any PPT $N$--qubit ($N>3$) state
acting on $\mathcal{S}((\mathbb{C}^{2})^{\ot N})$ with
$r(\varrho)\leq N$ is fully separable, i.e., it can be written as
\begin{equation}\label{fullySep}
\varrho=\sum_{i}p_{i}\varrho_{A_{1}}^{(i)}\ot\ldots\ot
\varrho_{A_{N}}^{(i)}.
\end{equation}
This implies immediately that we can restrict our considerations
to $r(\varrho)=5$. The same reasoning can be applied to
$\varrho^{\Gamma_{A}}$ and consequently
$r(\varrho^{\Gamma_{A}})\geq 5$. Finally, we can treat
$\sigma=\varrho^{\Gamma_{AB}}$ as a three--partite PPT state
acting on $\mathbb{C}^{2}\ot\mathbb{C}^{2}\ot\mathbb{C}^{3}$ (recall
that $\mathcal{S}(\mathbb{C}^{2}\ot\mathbb{C}^{2})$ is isomorphic to $\mathbb{C}^{3}$).
It follows from Ref. \cite{22N} that if $r(\varrho^{\Gamma_{AB}})\leq
3$ then $\sigma$ is a three--partite fully separable (see Eq.
(\ref{fullySep})) and thus, {\it via} reasoning from Ref.
\cite{Eckert} $\varrho$ is a four--party fully separable state.
Therefore we can assume that $r(\varrho^{\Gamma_{AB}})\geq 4$.

Let us now pass to the criterion. The Hermitian matrix we look for
has to be also symmetric as we need the condition $R(H)\subseteq
R(\varrho)$ to be satisfied. Consequently $r(H)\leq 5$ and we have
at most 25 parameters to determine. The condition for
$r(\varrho^{\Gamma_{A}})$ and $r(\varrho^{\Gamma_{AB}})$ under
which $H$ exists can be straightforwardly determined as in the
previous cases and reads
$[r(\varrho^{\Gamma_{A}})]^{2}+[r(\varrho^{\Gamma_{AB}})]^{2}>121$,
where one follows from the normalization condition. It implies
that there are no symmetric PPT extremal states with the following
triples of ranks $(5,7,9)$, $(5,8,8)$, $(5,8,9)$ (this agrees with
the fact that points being interiors points of two sets and not
the extremal point of the third one cannot be extremal points of
an intersection of these three sets). All this implies that we
have the following ranks for which it is possible that symmetric
four--qubit PPT entangled and extremal states exist:
$r(\varrho)=5$, $5\leq r(\varrho^{\Gamma_{A}})\leq 7$, and $4\leq
r(\varrho^{\Gamma_{AB}})\leq 8$.

Still, however, by virtue of Refs. \cite{2xN} and \cite{Eckert}
further simplifications can be inferred. For instance we can treat
$\sigma=\varrho^{\Gamma_{A}}$ as a $2\ot 6$ density matrix with
respect to the cut $B|ACD$. Now, if $\sigma$ is supported on
$\mathbb{C}^{2}\ot\mathbb{C}^{p}$ $(3< p\leq 6)$ then if follows
from Ref. \cite{2xN} that the condition
$r(\varrho^{\Gamma_{A}})>p$ has to be satisfied. Notice that for
$r(\varrho^{\Gamma_{A}})=p$, $\sigma$ is separable across $B|ACD$
and therefore $\varrho$ is fully separable. To see this explicitly
let us notice that $\varrho$ can be written as
$\varrho=\sum_{i}p_{i}\proj{e_{B}^{(i)}}\ot
\proj{\psi^{(i)}_{ACB}}^{\Gamma_{A}}$. Then, it suffices to
utilize the fact that tracing out the $A$ subsystem we get
three-qubit symmetric density matrix which is separable across the
$B|CD$ cut. Finally, utilizing similar approach as the one in Ref.
\cite{Eckert} one sees that $\varrho$ has to be fully separable.
Analogous arguments work also in the case of
$\varrho^{\Gamma_{AB}}$ with respect to e.g. $C|ABD$ cut.

In view of what was just said it seems then that generically the
only possible cases with respect to ranks
$(r(\varrho),r(\varrho^{\Gamma_{A}}),r(\varrho^{\Gamma_{AB}}))$
are $(5,7,7)$, and $(5,7,8)$.

\section{Conclusion}

Let us discuss shortly the presented results. Once more we need to
stress that in principle due to results of Gurvits \cite{Gurvits}
and Doherty \cite{Doherty1} the problem of separability in lower
dimensional Hilbert spaces can be regarded as solved. Still
however, it seems that some progress can be achieved in systems
like $3\ot 3$ or $2\ot 4$. Here one knows that except the states
that are detected by the transposition map (NPT states) there are
also entangled states with positive partial transposition. As we
do not have any unique structural criterion that could detect
these PPT states, an attempt to learn about the geometry of the
set of PPT states in the lower dimensional systems like $2\ot 4$
or $3\ot 3$ seems interesting. Some progress in this direction in
$3\ot 3$ systems has recently been obtained in
Refs.~\cite{HaKyePark,Ha2,Lieven,Ha2007}, where new classes of the
edge states have been found for all possible configurations of
ranks of a density matrix and its partial transposition. What is
even more important examples of these states with bi--ranks
$(4,4)$, $(5,5)$, and $(6,6)$ were shown to be extremal in the set
of PPT states in Refs. \cite{HaKyePark,Kim,Ha}.

The main purpose of the paper was to study extremality in the
convex set of PPT states in qubit--ququart systems and the main
purpose was to provide a operational criterion allowing to judge
if a given $\varrho$ is extremal. At the proof stage we learned,
however, about Ref. \cite{Leinaas} (see also Ref. \cite{Ha}) in
which similar criterion has already been given. In this work we
have provided our formulation of the criterion. It reduces the
question of extremality to the problem of solving a system of
linear equations. Using this approach we have reconstructed the
known literature results, i.e., that there are no extremal states
in $\mathcal{D}_{2,4}^{\mathrm{PPT}}$ ranks of which satisfy
$r(\varrho)+r(\varrho^{\Gamma_{A}})\geq 12$ (except for the case
of $r(\varrho)=r(\varrho^{\Gamma_{A}})=6$). We have investigated
the remaining cases of the bi--rank $(5,5)$, $(5,6)$, and $(6,6)$.
In the case of $(6,6)$ the criterion shows that there are no
extremal states. In the case of bi--rank $(5,5)$ we proved that
all the $(5,5)$ edge states are extremal in
$\mathcal{D}_{2,4}^{\mathrm{PPT}}$ implying that the famous
Horodecki PPT states \cite{PHPLA} are extremal. On the other hand,
existence of PPT entangled extremal states of bi--rank $(5,6)$ has
been confirmed numerically in Ref. \cite{Leinaas}, however,
without giving explicit examples. It means that in the $2\ot 4$
systems only the PPT entangled states of bi--rank $(5,5)$ and
$(5,6)$ can be extremal. We have also provided a class of states
that for some parameter region are $(5,6)$ edge states, while for
some other parameter region they constitute other examples of
$(5,5)$ extremal states.

Using the criterion, we have made an attempt to study the
structure of $(5,6)$ PPT states. For any $(5,6)$ $\varrho$ we
provided such a Hermitian $H$ that $R(H)\subseteq R(\varrho)$,
while the condition $R(H^{\Gamma_{A}})\subseteq
R(\varrho^{\Gamma_{A}})$ holds if some corresponding set of linear
equations has a solution. We asked what are the conditions imposed
on the matrix $B$ (see Eq.~(\ref{2x4})) if the solutions are of
some particular form. It is shown that whenever $B$ satisfies the
corresponding conditions the solution to the system of equations
exists.

Finally, we extended our considerations on extremality to the
simple systems of many-qubits as general three-qubit and
four-qubit symmetric PPT states. In the first case we have shown
that the only cases with respect to the respective ranks in which
one may expect extremal PPTES are $(4,4,4)$ and $(4,4,5)$.

\acknowledgments We are grateful to J. M. Leinaas for bringing our
attention to Ref. \cite{Leinaas}. R. A. acknowledges discussion with J. Stasi\'nska.
We acknowledge Spanish MEC/MINCIN projects TOQATA (FIS2008-00784) and QOIT (Consolider
Ingenio 2010), ESF/MEC project FERMIX (FIS2007-29996-E), EU
Integrated Project  SCALA, EU STREP project NAMEQUAM, ERC Advanced
Grant QUAGATUA, and Alexander von Humboldt Foundation Senior
Research Prize.

\end{document}